# TiN FILMS DEPOSITED BY LASER CVD: A GROWTH KINETICS STUDY

A. J. Silvestre and O. Conde

*Department of Physics, University of Lisbon Ed. C1, Campo Grande, 1700 Lisboa, Portugal*

**Abstract**

Results on the chemical composition, structure and growth kinetics of titanium nitride (TiN) films deposited on mild steel substrates by pyrolytic laser-induced chemical vapour deposition (LCVD) are presented. Golden coloured lines of TiN were deposited from a reactive gas mixture of $TiCl_4$, $N_2$ and $H_2$ using a continuous wave $TEM_{00}$ $CO_2$ laser beam as heat source. The chemical composition and structure of the films were determined by electron probe microanalysis (EPMA) and glancing incidence X-ray diffraction (GIXRD). A non-contact laser profilometer was used to measure the thickness profiles of the films. Using the data obtained in the steady-state region of the TiN laser-written lines, growth rates in the range 3.7 to 6.9 $\mu m.s^{-1}$ were deduced. The Arrhenius relation between the deposition rate and the deposition temperature yields an apparent activation energy of $46.9 \pm 3.8$ $kJ.mol^{-1}$. This result enabled us to conclude that under our deposition conditions the LCVD of TiN is controlled by mass transport in the vapour phase.

*Keywords:* Coatings; Titanium nitride (TiN); Laser chemical vapour deposition (LCVD); Growth mechanisms.

## 1. Introduction

Since the first report in 1972 [1], laser-induced chemical vapour deposition (LCVD) has been extensively studied as a powerful technique for selective-area deposition of thin films. The LCVD process is based upon the interaction of a laser beam with a reactive gaseous/vapour atmosphere and a substrate material. Film deposition occurs either by direct bond breaking in the precursor molecules due to resonant absorption of the laser radiation (photolytic LCVD) or by thermal decomposition of the reactant molecules at or near the surface of the laser heated substrate (pyrolytic LCVD). The extremely high deposition rates attained [2] associated with the highly localised heat source make LCVD an attractive technique for the deposition of a wide variety of useful materials [3], in different fields such as microelectronics and micromechanics.

Among the materials deposited by pyrolytic LCVD, the synthesis of titanium nitride (TiN) has been studied with particular interest due to the exceptional combination of physical and chemical properties of TiN [4] which makes it an excellent material for a wide variety of technological applications. Since 1990, several studies on the deposition of this ceramic compound by thermal LCVD have been published [5-13]. In most cases, a stationary laser beam has been employed yielding spot-shaped deposits, while only two studies on the laser direct writing of TiN stripes have been reported [10,13]. Despite the work already done, the LCVD of TiN is a field of investigation in which fundamental aspects are not yet fully understood, in particular chemical kinetics and growth mechanisms.

This paper reports on the chemical composition, structure and growth kinetics of laser direct writing of TiN lines deposited on mild steel substrates. The TiN lines were processed using a $CO_2$ laser and a gaseous mixture composed of $TiCl_4$, $N_2$ and $H_2$.

## 2. Experimental procedure

Details of the deposition system and of the procedure used for LCVD of TiN have been presented elsewhere [9,10] and only a brief description is given here. The experiments were conducted in a closed stainless steel reactor provided with a ZnSe window for transmission of the laser beam at perpendicular incidence to the substrate. A cw $CO_2$ laser operating predominantly in $TEM_{00}$ mode was used. The reaction chamber was always evacuated to a base pressure of $6.7 \times 10^{-4}$ Pa before the introduction of the reactant gas mixture. Computer-controlled XY stages enabled the substrates to be scanned under the laser beam. Square samples of mild steel (5 cm wide) finished with 1 μm diamond spray were used as substrates. Prior to their insertion into the reactor, the substrates were ultrasonically cleaned in acetone.

*2.1. Deposition conditions*

The films were deposited from a static gaseous mixture of $TiCl_4$, $N_2$ and $H_2$ at total pressure of $2.8 \times 10^4$ Pa. In all the experiments the partial pressure of $TiCl_4$ was $9.3 \times 10^2$ Pa and for $N_2$ and $H_2$ the same partial pressure of $133.3 \times 10^2$ Pa was used. The laser spot diameter at the substrate surface was 1.71 mm, the incident laser power was varied between 400 and 700 W, and the laser power density or irradiance thus took values between $1.51\times10^4$ and $2.64\times10^4$ W.cm$^{-2}$ [9]. The scanning velocity of the substrates relative to the laser beam was varied between 0.5 and 5.0 mm.s$^{-1}$.

The deposition process is characterised by the existence of a threshold irradiance of $1.5\times10^4$ W.cm$^{-2}$ below which the reaction temperature is not attained, and for laser irradiance values higher than $2.26\times10^4$ W.cm$^{-2}$, melting of the film/substrate was observed. The best results were obtained for laser irradiances in the range $1.88\times10^4$ to $2.26\times10^4$ W.cm$^{-2}$ and scanning velocities between 2.0 and 4.0 mm.s$^{-1}$ [10].

*2.2. Characterisation*

The chemical composition of the TiN laser-written lines was studied using an automated SX50 Cameca electron probe microanalyser (EPMA). Atomic concentrations were determined by wave dispersive spectrometry (WDS) applying the classical standardisation technique. Pure stoichiometric TiN was set as a standard for evaluation of nitrogen content while pure Ti was used for titanium content. The composition was corrected using the PAP program [14]. The structure of the coatings was determined by X-ray diffraction using a Siemens D5000 diffractometer. All the films were analysed at a glancing angle of 1º with Cu-K$\alpha$ radiation. The deposition rates were estimated based on the thickness profiles of the coatings, measured with a Rodenstock RM600 non-contact laser profilometer.

**3. Results and discussion**

All the lines deposited within the selected parameters exhibit the characteristic TiN golden colour, good adherence and a broad Gaussian-type profile (fig. 1). As reported previously [10], these films also show a very fine and uniform equiaxed grain structure, only resolved by SEM at high magnification ($\times15000$).

*3.1. Chemical composition*

Since stoichiometry determine the main tribological, electrical and optical properties of TiN coatings [4], an investigation of the chemical composition of the synthesised material was carried out. Results of EPMA revealed that the deposited titanium nitride is nearly stoichiometric. The atomic ratio [N]/[Ti] measured at different points along a line perpendicular to the scanning direction of the laser beam is plotted in figure 1. From this figure it can be concluded that the chemical composition does not vary from the centre to the periphery of the line. This result agrees with the SEM observations that revealed no variation in the type of microstructure along the transverse direction of the line [10]. A similar behaviour was found for all the films deposited with the above mentioned selected parameters.

By contrast, there is an evolution of the atomic ratio [N]/[Ti] along the central line of the films, in the scanning direction. A representative example of the results obtained can be observed in figure 2, where the atomic ratio [N]/[Ti] is plotted as a function of the distance to the origin of the laser-substrate interaction for a film deposited with an irradiance of $2.26 \times 10^4$ W.cm$^{-2}$ and a scanning velocity of 4.0 mm.s$^{-1}$. In the steady-state regime the stoichiometry of the deposited material is approximately constant and slightly overstoichiometric. However, in the transient region of the deposition process the films are clearly understoichiometric ([N]/[Ti]<1). The lack of nitrogen detected in this region is probably due to the fact that in the early stages of the deposition process the surface temperature induced by the laser beam is not sufficient to dissociate molecular nitrogen in the required proportion. It should be noted that while progressing in the transient region towards the stationary regime of deposition, nitrogen concentration increases with the increase in temperature.

Figure 3 displays the atomic ratio [N]/[Ti] as a function of the scanning velocity for laser irradiances of $1.88 \times 10^4$ and $2.26 \times 10^4$ W.cm$^{-2}$. Each data point represents a mean value of the [N]/[Ti] ratios measured at successive positions, spaced 1 mm along the centre of each TiN line. As can be seen, chemical composition does not vary either with laser irradiance or with scanning velocity. This means that the deposition temperature range implied by the values of these two processing parameters has no detectable effect on the stoichiometry of the films.

*3.2. Structural analysis*

A typical GIXRD pattern of the laser deposited material can be observed in figure 4. The diffractograms obtained are all matched by δ-TiN JCPDS card no. 38-1420, confirming that only one crystalline phase was formed during the LCVD process. The relative intensity of the peaks measured in the diffraction patterns agrees with the standard relative intensities of δ-TiN with the grains randomly oriented, implying that the synthetized material is non-textured. Following the procedure suggested by Cullity [15], the lattice parameter $a_o$ was determined from the intercept value of the linear plot of $\cos^2\theta / \sin\theta$ *vs*. $a(\theta)$, where $a(\theta)$ represents the reticular parameter estimated from each single peak centred in the angular position $2\theta$ (fig. 5). The lattice parameter of the deposited TiN is very close to the standard value $a_o = 2.24$ Å of stoichiometric δ-TiN, in agreement with the EPMA data.

*3.3. Growth kinetics*

Although a pyrolytic LCVD process may present a more complex structure than conventional CVD, mass transport in the vapour phase and surface kinetics are still the most important rate-limiting mechanism in deposition. Arrhenius diagrams, i.e. the logarithm of the deposition rate *versus* the reciprocal temperature curves, are a common convenient way to infer these reaction control mechanisms. If the rate-limiting step during an LCVD process is the surface chemical reaction, the activation energy, deduced from the slope of the Arrhenius plots, is usually a few hundred $kJ.mol^{-1}$. By contrast, when the deposition process is controlled by mass transport in the vapour phase the growth rate is weakly dependent on temperature, yielding a low value for the activation energy.

Since all the TiN lines present a broad Gaussian-type profile, we have chosen the central film thickness values as a criterion to evaluate the deposition rates. Figure 6 shows these values as a function of the scanning velocity for different laser irradiances. The thicknesses were measured in the steady-state region [10] of the laser-written TiN lines. As depicted, the film thickness decreases when the scanning velocity increases and, at constant scanning velocity, it increases with laser irradiance. This behaviour was expected since, for a given film/substrate system, the scanning velocity and laser irradiance are the processing parameters which most strongly determine the maximum temperature attained during the laser-material interaction.

The Arrhenius plot represented in figure 7 shows the logarithm of the deposition rate as a function of the reciprocal central deposition temperature. The deposition rates were calculated as the ratio of central film thickness to interaction time, which can be approximated to the ratio of the beam diameter to the scanning velocity. Deposition rates between 3.7 and 6.9 $\mu m.s^{-1}$ (table 1) were calculated, which are two orders of magnitude higher than deposition rates reported for CVD of TiN [16]. The central deposition temperatures were calculated by using the three-dimensional transient heat transfer model for uniformly moving finite slabs developed by Kar and Mazumder [17]. It can be seen from the Arrhenius diagram that only one straight line is needed to fit the data. Thus, within the deposition temperature range investigated and for the reactive atmosphere considered, only one mechanism is responsible for the TiN film growth. The slope of the straight line yields an apparent activation energy of $46.9 \pm 3.8$ $kJ.mol^{-1}$. This value is in very good agreement with those reported, from 24 to 55 $kJ.mol^{-1}$, for conventional CVD of TiN grown from the same reactive mixture as used in this work, for the case where the deposition process is limited by diffusion of the gaseous species [18]. We can thus conclude that, under the experimental conditions used in this study, mass transport of the reactive gaseous species is the rate limiting step for titanium nitride film growth. In contrast, Azer reported [19] a higher activation energy of 65.2 $kJ.mol^{-1}$ for spot-shaped films of TiN deposited by LCVD on Ti-6Al-4V substrates, between 1400 and 1650 K, which could be mainly due to the poor statistics of the data used in Arrhenius fitting. Nevertheless, the value deduced by Azer is still in the range of activation energy values accepted as characteristic of growth mechanisms controlled by mass transport.

The equiaxed small grain size microstructure uniformly developed in these films is related to the controlled growth mechanisms described, in contrast with previously observed evolving microstructures [9] due to transitions from surface kinetics at lower temperatures to mass transport mechanisms at higher temperatures.

## 4. Conclusions

Golden coloured laser-written TiN lines with broad Gaussian thickness profiles were deposited on mild steel by pyrolytic LCVD using a cw $TEM_{00}$ $CO_2$ laser. The coatings are nearly stoichiometric and were deposited as polycrystalline non-textured material with a lattice parameter close to the standard value for stoichiometric δ-TiN.

Based on the thickness profiles of the films, growth rates in the range 3.7 to 6.9 $\mu m.s^{-1}$ were deduced. The Arrhenius equation for the deposition rate as a function of the deposition temperature yields an apparent activation energy of 46.9 ± 3.8 kJ.mol$^{-1}$. By comparing this result with the CVD values it can be concluded that under the deposition conditions used in this work the LCVD of TiN is controlled by the diffusion of the gaseous species into the reaction zone.

**Acknowledgments**

The authors great fully acknowledge the financial support of Junta Nacional de Investigação Científica e Tecnológica (contract no. PMCT/C/MPF/470/90) and PhD grant for AJS.

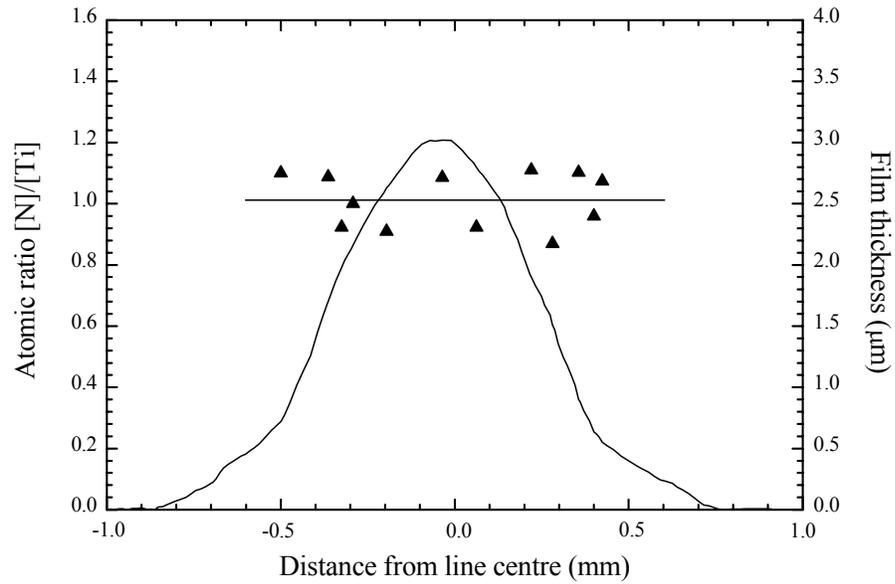

**Figure 1** - Thickness profile and chemical composition measured at the same cross-section of a TiN line deposited at I = 2.26×10$^4$ W.cm$^{-2}$ and v = 4.0 mm.s$^{-1}$. ▲, [N]/[Ti] atomic ratio as determined by EPMA; —, mean value of [N]/[Ti] atomic ratio (1.01).

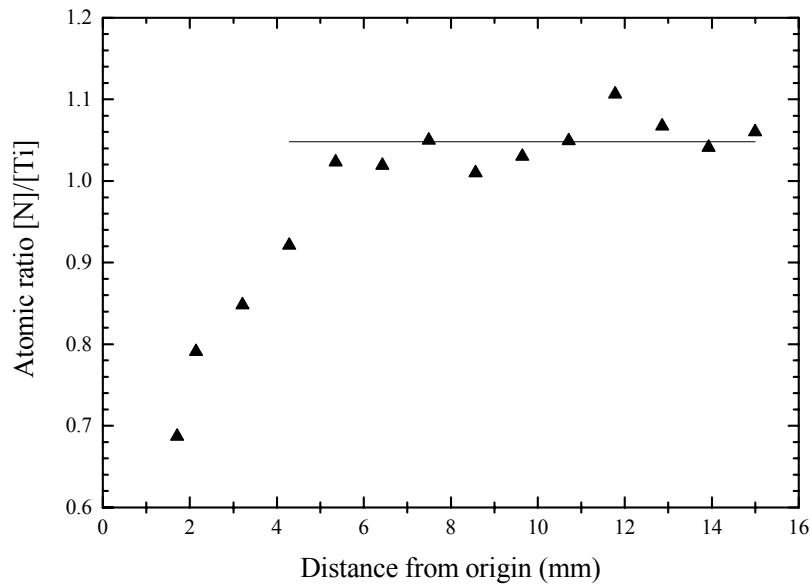

**Figure 2** - [N]/[Ti] atomic ratio *versus* distance from origin of a TiN line deposited at I = 2.26×10$^4$ W.cm$^{-2}$ and v = 4.0 mm.s$^{-1}$. ▲, [N]/[Ti] atomic ratio as determined by EPMA; —, mean value of [N]/[Ti] atomic ratio (1.05) in the stationary region of the deposition process.

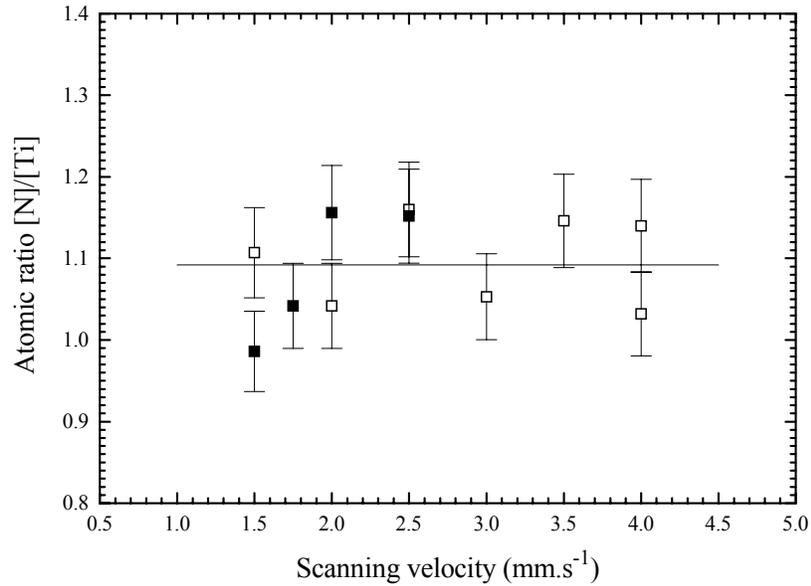

**Figure 3** - EPMA results for the chemical composition of the films *versus* the scanning velocity, for two different laser irradiance values: ■, I = 2.26×10$^4$ W.cm$^{-2}$; □, I = 1.88×10$^4$ W.cm$^{-2}$. The full line gives the mean value of the experimental data (1.09).

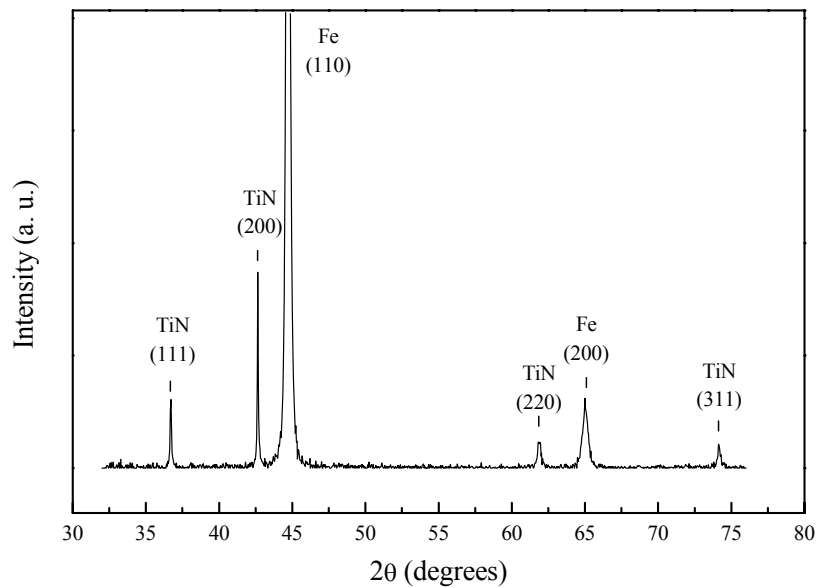

**Figure 4** - X-ray spectra of a TiN line deposited at I = 1.88×10$^4$ W.cm$^{-2}$ and v = 4.0 mm.s$^{-1}$.

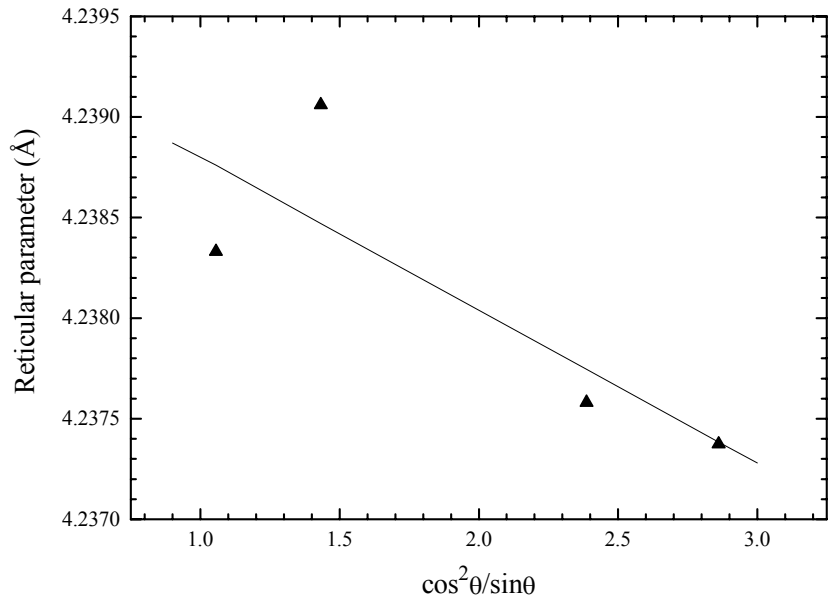

**Figure 5** - Plot a($\theta$) *versus* $\cos^2\theta/\sin\theta$ corresponding to the X-ray spectra shown in figure 4. The intercept value of the linear fitting gives $a_o$ = 4.2396 Å.

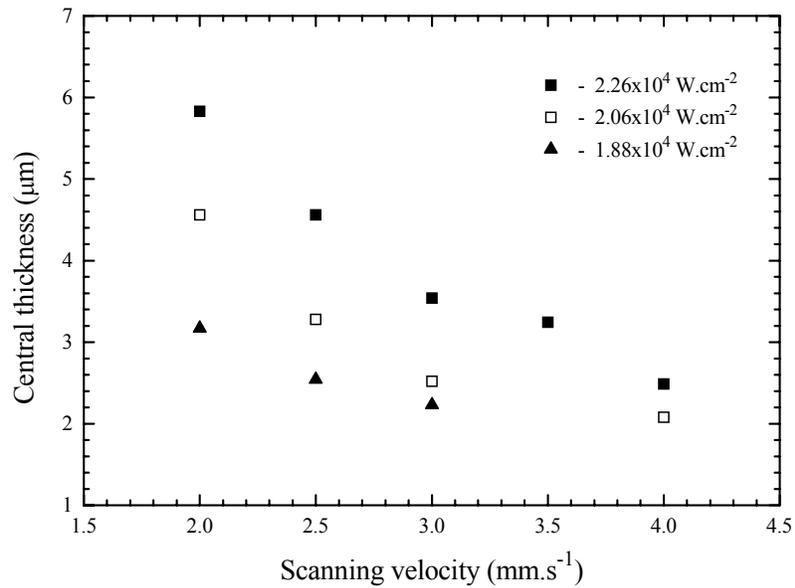

**Figure 6** - Central film thickness as a function of the scanning velocity for diferent laser irradiances.

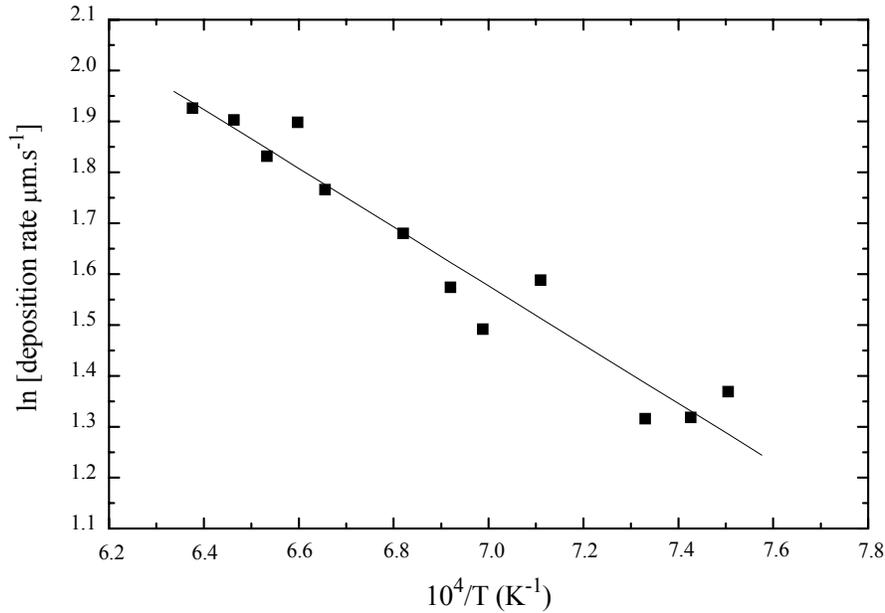

**Figure 7** - Logarithm of the apparent deposition rate as a function of the reciprocal central deposition. ■, measured values; —, line fitting.

**Table1** - Experimental data used to construct the Arrhenius plot.

| I (W.cm$^{-2}$) | v (mm.s$^{-1}$) | $t_{int}$ (s) | T (K) | th (μm) | $r_{TiN}$ (μm.s$^{-1}$) |
|---|---|---|---|---|---|
| | 2.0 | 0.850 | 1568.3 | 5.83 ± 0.09 | 6.86 |
| | 2.5 | 0.680 | 1547.2 | 4.56 ± 0.28 | 6.71 |
| 2.26×10$^4$ | 3.0 | 0.567 | 1530.8 | 3.54 ± 0.14 | 6.24 |
| | 3.5 | 0.486 | 1515.5 | 3.24 ± 0.19 | 6.67 |
| | 4.0 | 0.425 | 1502.6 | 2.48 ± 0.27 | 5.85 |
| | 2.0 | 0.850 | 1466.3 | 4.56 ± 0.21 | 5.36 |
| 2.07×10$^4$ | 2.5 | 0.680 | 1445.1 | 3.28 ± 0.08 | 4.82 |
| | 3.0 | 0.567 | 1431.1 | 2.52 ± 0.17 | 4.44 |
| | 4.0 | 0.425 | 1406.4 | 2.08 ± 0.09 | 4.89 |
| | 2.0 | 0.850 | 1364.2 | 3.17 ± 0.14 | 3.73 |
| 1.88×10$^4$ | 2.5 | 0.680 | 1346.6 | 2.54 ± 0.27 | 3.74 |
| | 3.0 | 0.567 | 1332.5 | 2.23 ± 0.15 | 3.93 |